\documentclass[12pt,preprint]{aastex}

\begin{document}

\slugcomment{Accepted for publiction in the Astrophysical Journal}

\title{Cool Companions to White Dwarfs from the 2MASS Second Incremental 
Data Release} 
\author{Stefanie Wachter}  
\affil{SIRTF Science Center, California Institute of Technology, 
MS 220-6, Pasadena, CA 91125}
\email{wachter@ipac.caltech.edu}

\author{D. W. Hoard}
\affil{SIRTF Science Center, California Institute of Technology, 
MS 220-6, Pasadena, CA 91125}
\email{hoard@ipac.caltech.edu}

\and 

\author{Kathryn H. Hansen, Rebecca E. Wilcox, Hilda M. Taylor, 
Steven L. Finkelstein }
\affil{University of Washington, Department of Astronomy, Box 351580, 
Seattle, WA 98195}

\begin{abstract}
We present near-infrared magnitudes for all white dwarfs (selected from the 
catalog of McCook \& Sion) contained in the 2 Micron 
All Sky Survey Second Incremental Data Release. 
We show that the near-IR color-color diagram is an effective
means of identifying candidate binary stars containing a WD and a
low mass main sequence star.  The loci of single WDs and WD + red
dwarf binaries occupy distinct regions of the near-IR color-color
diagram.  We recovered all
known unresolved WD + red dwarf binaries located in the 2IDR sky
coverage, and also identified as many new candidate binaries
(47 new candidates out of 95 total). 
Using observational near-IR data for WDs and M--L dwarfs, we have
compared a sample of simulated WD + red dwarf binaries with our
2MASS data.  The colors of the simulated binaries are dominated
by the low mass companion through the late-M to early-L spectral
types.  As the spectral type of the companion becomes progressively
later, however, the colors of unresolved binaries become
progressively bluer.  Binaries containing the lowest mass
companions will be difficult to distinguish from 
single WDs solely on the basis of their near-IR colors. 
\end{abstract}

\keywords{Binaries: general --- infrared: stars ---
          stars: fundamental parameters, surveys --- white dwarfs}

\section{Introduction}

In the search for extrasolar planets, various methods have been 
employed to detect the signatures of faint stellar and sub-stellar 
companions.  For main sequence primary stars, faint low mass 
companions are often hidden in the glare of the more luminous 
primary, and radial velocity variations are small and therefore 
difficult to detect.  On the other hand, observing low mass 
companions to white dwarfs (WDs) offers many advantages compared 
to main sequence primaries.  Since WDs are less luminous than 
main sequence stars, the brightness contrast compared to a 
potential faint companion is significantly reduced. Most 
importantly, the markedly different spectral energy distributions 
of the WDs and their low mass companions makes the detection and 
separation of the two components relatively straightforward even 
with simple broad-band multi-color photometry.

Because WDs have traditionally been identified and studied via 
observations in the blue part of the spectrum, comparatively little 
is known about their infrared (IR) properties.  The recent 
discovery that very cool WDs are much bluer in the IR than 
previously thought to be the case \citep{hodgkin00} highlights 
how little is known about WD spectral properties at longer 
wavelengths.  Consequently, we are studying the group near-IR 
photometric properties of WDs. In this paper, we present analysis 
of the near-IR color-color diagram of WDs, which demonstrates a 
means of identifying candidates for WDs with close (unresolved), 
cool, low mass stellar or sub-stellar companions.

\section{Target Selection and Identification}

We selected the WDs in our sample from the catalog of 
spectroscopically identified WDs by \citet[][hereafter, MS99]{ms99}.  
We extracted all WDs from MS99 that are contained in the 
sky coverage of the 2MASS Second Incremental Data Release 
(2IDR; e.g., \citealt{skrut95,skrut97})\footnote{Also see 
\url{http://pegasus.phast.umass.edu/}.}.  
Due to potentially large and often unknown proper motions, 
and other uncertainties in published positions, we first 
identified each WD in optical images from the Digitized 
Sky Survey (DSS). The WD in the optical image was then 
matched with sources in the 2MASS 2IDR images and point 
source catalog. Our identification of the optical counterpart 
was based on published finding charts whenever possible; 
for example, using the charts in the LHS atlas \citep{luyten79}, 
the Giclas proper motion survey and lists of suspected WDs 
(e.g., \citealt{G58} through \citealt{G80}), and the 
Montreal-Cambridge-Tololo survey \citep{lamontagne00}, to name 
only a few sources.  The WDs for which no finding chart 
could be located in the literature were identified from a 
combination of published coordinates, proper motion, and color 
in the DSS images. A catalog detailing accurate J2000 positions 
together with references to individual finding charts and 
our method(s) of identification will be presented in a future paper.

\subsection{Number Statistics}

MS99 list 2249 spectroscopically identified WDs, 1235 of which 
are located in the sky coverage of the 2MASS 2IDR.  For 47 WDs, 
we could not (re)establish an optical identification.  
This was mainly due to insufficient accuracy in the published 
finding charts and/or coordinates that made it impossible to 
decide with confidence between several stars close to the 
given positions.  In some cases, WDs listed in MS99 have 
subsequently been reclassified as AGN, Seyfert galaxies, or 
hot subdwarfs.  A few WDs appear in MS99 multiple times under 
different designations.  For 27 WDs, no IR magnitudes could 
be obtained from the 2MASS 2IDR point source catalog despite 
having an identified optical counterpart.  In the majority of 
cases, this is due to blending of the WD with unrelated field 
stars in the 2MASS images.  Detailed comments on particularly 
problematic identifications will be provided in a future paper. 

The 2MASS completeness limits (defined by photometry with 
signal-to-noise of S/N $>$ 10) are $J = 15.8$, $H = 15.1$, 
and $K_{\rm s}=14.3$.  The survey detection limits are approximately 
one magnitude fainter in each band.  Of the 1161 WDs for which 
we could establish secure identifications, 759 are detected in 
the 2MASS 2IDR with varying degrees of accuracy.  The remaining 
402 WDs are undetected, meaning that while we have securely 
identified an IR counterpart for these WDs, there is no 
corresponding entry in the 2MASS 2IDR point source catalog.  
Many of the formally undetected WDs appear to lie just below 
the detection limits of the survey, as faint objects are 
often visible in the 2MASS images at the correct positions.

\section{Analysis and Discussion}

\subsection{The IR Color-Color Diagram}
\label{s-wdccd}

\begin{figure}[tb]
\epsscale{1.00}
\plotone{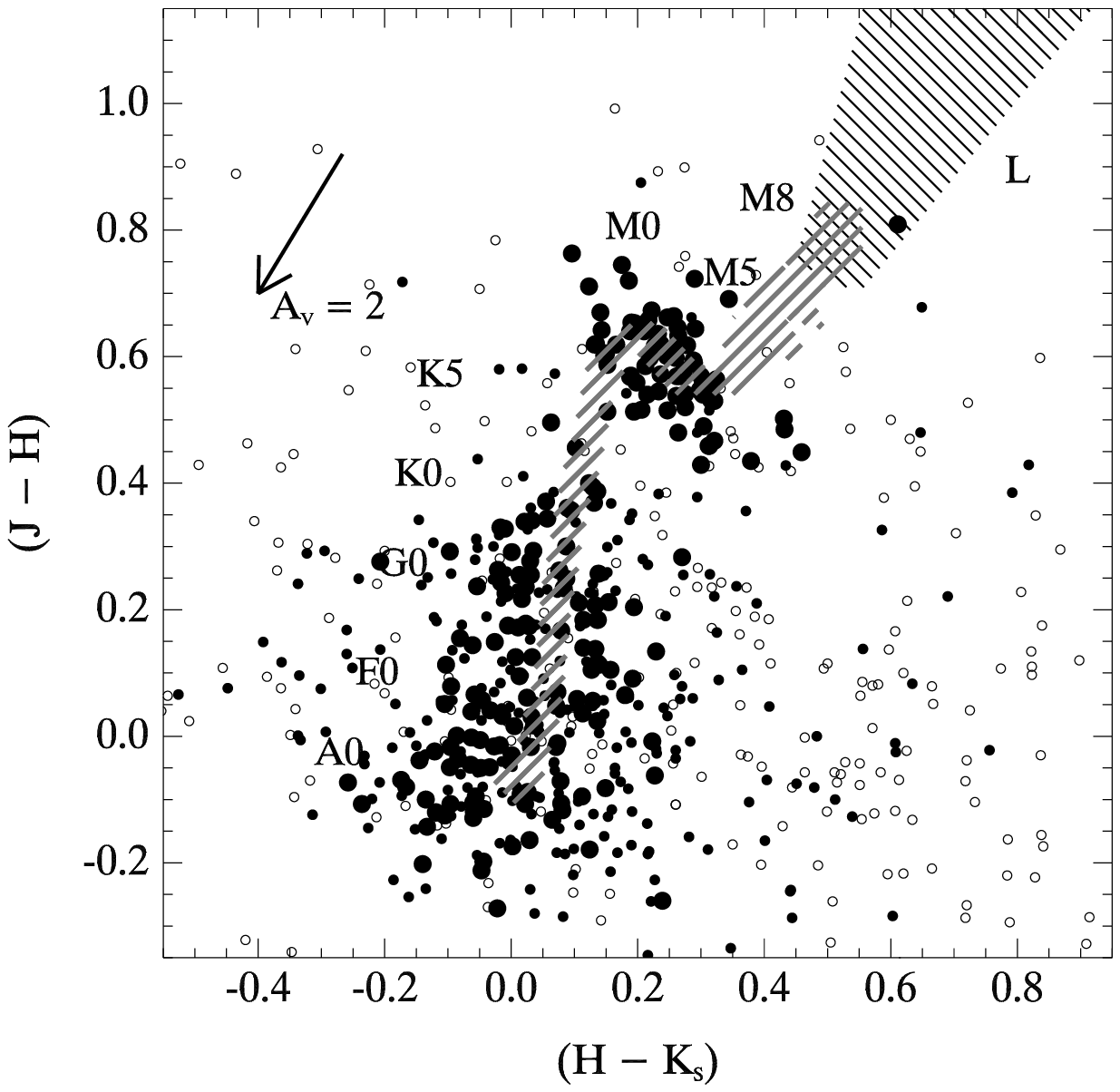}
\epsscale{1.00}
\figcaption{Near-IR color-color diagram for WDs from MS99 that 
are detected in the 2MASS 2IDR.  Also shown are the fiducial 
tracks for the main sequence (/// cross-hatches) and the region 
occupied by L dwarfs ($\backslash\backslash\backslash$ cross-hatches).   
The points are symbol-coded according to the $1\sigma$ 
uncertainties of the photometry (see section \ref{s-wdccd}). 
\label{f-ccdWD} }
\end{figure}

Figure \ref{f-ccdWD} illustrates the results of our study as 
a near-IR color-color diagram of WDs. We have plotted all WDs 
detected in the 2MASS 2IDR, together with the fiducial tracks 
of the main sequence and the region occupied by L dwarfs.  
The positions of the spectral type labels are offset 
horizontally for A0--K5, and vertically for M0--M8. 
The main sequence data up to M5 were taken from \citet{bb88} 
and transformed to the 2MASS photometric system using the 
relations in \citet{carpenter01}, while the colors for 
late-M and L dwarfs represent mean 2MASS observational data 
from \citet{gizis00} and \citet{kirk00}.  The points are 
symbol-coded according to the 1-$\sigma$ uncertainties of 
the original IR magnitudes:\ large filled circles = $\sigma<0.1$ mag 
for $J$, $H$, and $K_{\rm s}$; small filled circles = $\sigma>0.1$ mag 
for at least one of $J$, $H$, or $K_{\rm s}$; small unfilled 
circles = at least one magnitude is close to the 2MASS faint 
detection limit and lacks a formal uncertainty.  

If we examine only the data points with the smallest uncertainties 
(large filled circles), then our color-color diagram exhibits 
two prominent concentrations of points.  One group is clustered 
around the main sequence track of early spectral types to 
about K0, and another group is clustered around the locus of 
main sequence M stars. We expect that the former group contains 
isolated WDs and the WD components of wide 
(resolved\footnote{During our study, we found that, in general, 
binaries with separations of $d \leq 2\arcsec$ are unresolved in 
the 2MASS images, while those with separations of 
$d \geq 4\arcsec$ are resolved.  We assessed known binaries with 
separations of $2\arcsec < d < 4\arcsec$ on a case-by-case basis.}) 
binaries.  
The latter group contains close (unresolved) binaries 
consisting of a WD and a low mass main sequence companion, in 
which the red spectral energy distribution of the companion 
dominates the overall color. 

\begin{figure}
\epsscale{1.0}
\plotone{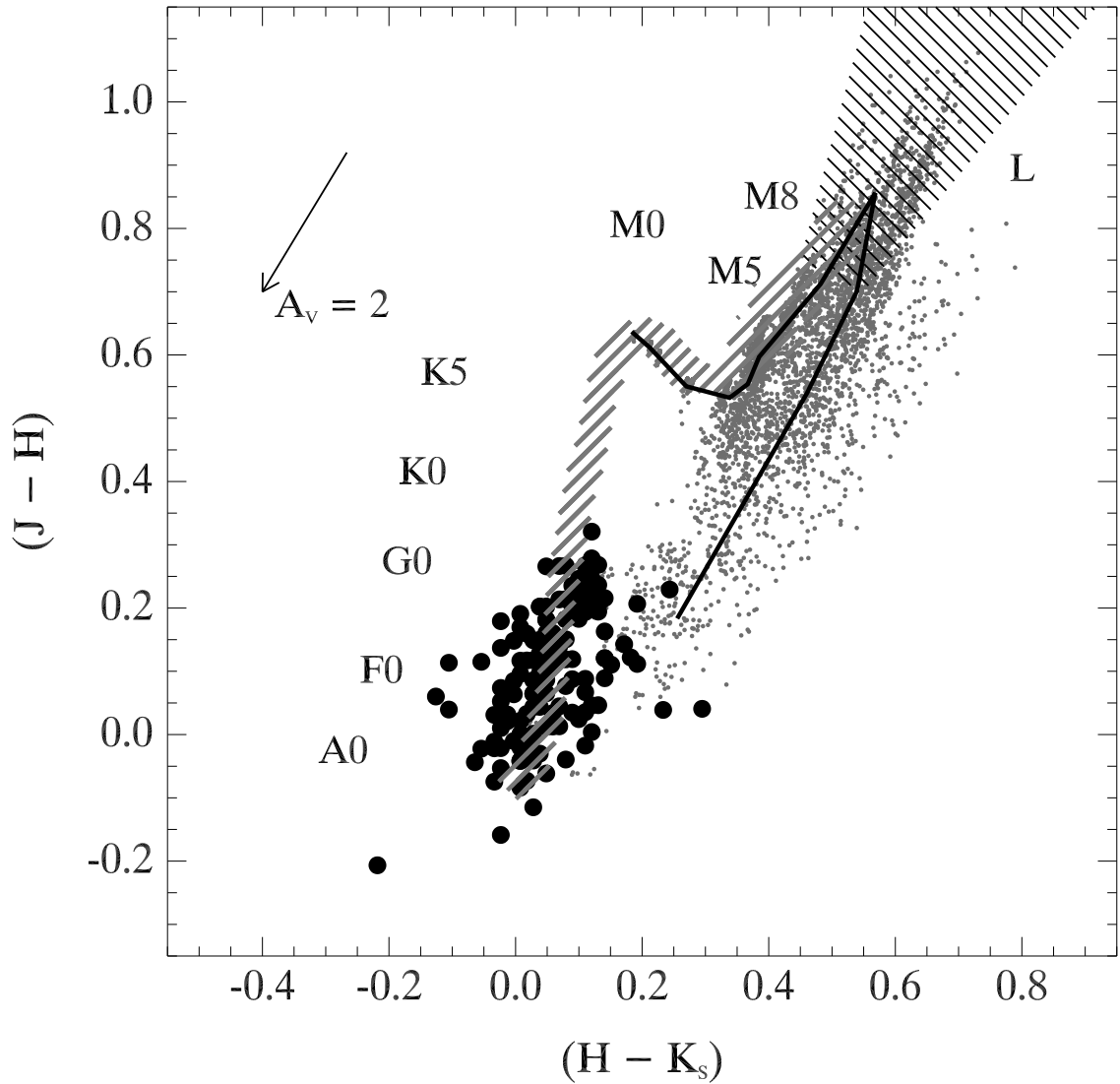}
\epsscale{1.0}
\caption{As in Figure \ref{f-ccdWD}, but showing the single, 
cool WD data from \cite{berg01} (large black circles) and 
simulated WD + main sequence star (M--L) binaries (small grey 
circles; see Section \ref{s-simbins}).  The solid black line 
is a schematic representation of the displacement in the 
color-color diagram caused by combining a given WD with a 
successively later spectral type companion.
\label{f-simWD} }
\end{figure}

In Figure \ref{f-simWD}, we show the near-IR color-color diagram 
of 152 {\it single}, cool WDs (large black circles) from the 
study of \citet{berg01}.  The data have been transformed from 
the CIT to the 2MASS photometric system using the relations in 
\citet{carpenter01}.  Typical photometric uncertainties in the 
transformed Bergeron data are about 5\%, with a few 
objects having larger uncertainties on the order of 10\%.  It 
is clear from the Bergeron data that single WDs populate the 
near-IR color-color diagram close to the locus of A--G main sequence 
stars, corresponding to the first cluster of points in our 2MASS 
color-color diagram. Unresolved double degenerate (WD + WD) 
binaries would, of course, also be located in this region and cannot 
be distinguished from single WDs in the color-color diagram.  
In comparison to the Bergeron sample, our data (limited to the 
sub-set with photometric uncertainties of $\sigma_{\rm JHK} < 0.1$ mag) 
cover a somewhat larger range in color space. This is partially due 
to the fact that the Bergeron sample was
selected to include only cool (T$_{\rm eff} \lesssim 12,000$ K) 
WDs with known 
parallaxes, while our sample contains a significant number of hotter WDs. 
We performed a literature search that yielded temperatures 
for 114 WDs in our low-uncertainty sub-set, 55 of which 
have T$_{\rm eff} > 12,000$ K.
Those hot WDs generally populate the 
blue (lower left) corner of the color-color diagram around (and below) 
the locus of the main sequence A stars. 
Further differences between the color distributions of the two data sets
are due to the lack of an exact 1:1 match of the particular WDs contained 
in each sample (that is, due to individual color differences from one 
WD to another), combined with the uncertainties in 
the photometry. Even a relatively small uncertainty 
of $\lesssim \pm 0.1$ mag in each color allows for 
a substantial shift in the placement of an individual object in the 
color-color diagram.  The 206 WDs with the smallest uncertainties in 
our data set (large filled circles in Figure \ref{f-ccdWD}) have 
mean uncertainties of $\langle\sigma_{\rm H-K_{\rm s}}\rangle = \pm0.07$ 
mag and $\langle\sigma_{\rm J-H}\rangle = \pm0.06$ mag, while those in 
the Bergeron sample are on the order of $\pm 0.07$ mag in each color index.

As mentioned above, we identify the clustering around the M star 
fiducial track in Figure \ref{f-ccdWD} as unresolved binary systems 
containing a WD and a low mass main sequence companion. 
The gap between the two data clusters in Figure \ref{f-ccdWD} 
(coincident with the locus of K0--K5 main sequence stars) can
be attributed to several factors. As indicated by the Bergeron sample, we do 
not expect single WDs in this color region. Consequently, only a binary 
consisting of a WD and a K dwarf companion would be located in this area
of the color-color diagram. Such composite systems are difficult to 
identify since 
the K star overwhelms the combined spectrum at optical--near-IR wavelengths.  
Furthermore, 
such systems are intrinsically rare simply due to a mass function effect;
that is, if the 
binaries formed from a random pairing of stars from the same initial mass 
function, then there are, in general, fewer K stars than M stars 
as potential companions.  Assuming a standard initial mass function 
\citep{kroupa02}, 
we calculate 0.079 for the expected ratio of K0--K5 to M0--M5 stars.
This can be compared to the number of objects in the color bins 
corresponding 
to those spectral types in our data sample, for which we derive a ratio 
of $\approx0.07$.
However, we caution that the photometric uncertainties make it unclear 
whether some of the systems belong to the M or K spectral type bins. 
A change in the spectral type classification of these systems could 
alter the value of this ratio, between extreme cases 
of $\approx 0.06$--$0.20$.

Based on a comparison with the location 
of single WDs in the color-color diagram (from Figure \ref{f-simWD}), 
we selected all objects with $(J-H) > 0.4$ mag as WD + low mass main 
sequence star binary candidates.  After eliminating objects with 
the highest photometric uncertainties (small unfilled circles), we 
find 95 such binary star candidates. Thirty-nine (41\%) of these 
candidates are already listed as binaries in MS99.   However, the 
references to their binary status contained in MS99 reveal that 
four of these are wide (resolved) binaries, whereas our 2MASS 
photometry suggests that the WD component may also be an 
unresolved WD + red dwarf\footnote{Throughout this paper we will 
refer to both M and L type main sequence stars as ``red dwarfs.''} 
pair.  We were unable to locate published information about the 
separations of another five of the known binaries.  We also 
performed a literature search with SIMBAD for each of our 
candidates, which identified 13 additional known binaries that 
are not classified as such in MS99. Table \ref{t-bincans} lists 
all of our candidates together with notes and references regarding 
their binary classification status.  Altogether, approximately 
half (47 out of 95) of our candidates are previously {\it unknown} 
to be binaries. 

Because of the limited spatial resolution of 2MASS 
($2\arcsec$ pixel$^{-1}$), we investigated whether a chance 
superposition of a red field star could have produced a significant 
number of our binary candidates.  However, visual inspection of 
the optical (DSS) and IR images shows that none of our candidates 
are located in crowded fields, so that the likelihood of a chance 
superposition is very small.  There are 50 additional objects 
classified as binaries in MS99 that are detected in the 2MASS 2IDR, 
but were {\it not} selected by our color criterion as unresolved 
binary candidates.  Forty-two of those are known to be resolved 
binaries, in which we can separately detect the WD and red dwarf 
components.  Four are unresolved double degenerate (WD + WD) 
binaries, which are indistinguishable from single WDs in the 
color-color diagram.  The remaining four are unresolved WD + main 
sequence binaries in which the companion has spectral type earlier 
than K (hence, these systems are dominated by the bright companion 
and fall along the main sequence in the color-color diagram, 
intermingled with the single WDs and below our color selection 
criterion).  Thus, our 2MASS data allows us to recover {\em all} 
of the known, unresolved WD + red dwarf binaries from MS99 that 
are detected in the 2MASS 2IDR.

\subsection{Simulated Binary Colors}
\label{s-simbins}
 
While the red colors of the low mass companions provide a striking 
contrast to those of the WDs, the luminosity of the low mass stars 
also rapidly declines as the companion mass decreases.  In order to 
properly evaluate the relative contribution of the WD and the red 
companion to the overall color of a binary, we calculated the 
expected colors from random pairings of a WD and a M--L dwarf.  
We combined (as fluxes) the $JHK_{\rm s}$ magnitudes of the sub-set 
of Bergeron WDs with known distances (and, hence, absolute 
magnitudes) with the absolute $JHK_{\rm s}$ magnitudes of M--L 
stars \citep{hawley02} to produce a set of simulated binary colors.  
The resulting simulated binaries are shown as small grey circles in 
Figure \ref{f-simWD}. Recall that the large black circles in this 
figure represent the original single WD data from \citet{berg01}.  
The solid black line is a schematic track demonstrating the effect 
on the combined color as a given WD is successively combined with 
later and later spectral type stars.  In general, after a short 
excursion into the region occupied by early-L dwarfs, the red 
companion becomes too faint to dominate the combined colors of the 
system. For progressively later L type companions, the binary color 
moves blueward, back towards the locus of single WDs. 

It is apparent from this simulation that some systems with colors 
close to those of single WDs may actually contain low mass L dwarf 
companions.  Consequently, we selected a second group of WDs from 
our 2MASS data set that satisfy the color criteria 
$0.2 \leq (H-K_{\rm s}) \leq 0.5$ and $0.1 \leq (J-H) \leq 0.4$.  
These 15 objects are listed as tentative binary candidates in 
Table \ref{t-tentbincans}.  Four of them are identified as known 
binaries by MS99.  Interestingly, however, all of these are 
classified as wide (resolved) binaries, which may imply that 
they are really triple systems, in which the WD component is 
actually an unresolved WD + red dwarf binary as well.  
Another one of them (WD0710+741) is a known close (unresolved) 
binary.  \citet{marsh96} used radial velocity data to estimate 
the mass of the WD's companion as 0.08--0.10M$_{\odot}$, which 
is consistent with a late-M to early-L spectral type.

\subsection{Comparison to Previous Studies}

Previous dedicated searches for cool companions to WDs using near-IR 
observations have been conducted by \citet[][henceforth, P83]{probst83}, 
\citet{zuckerman92}, and \citet[][henceforth, GAN00]{green00}.  
P83 surveyed 113 relatively bright WDs in $K$ with a 
12\arcsec-aperture, single-pixel InSb detector.  
This survey was somewhat hampered by the lack of spatial resolution 
and because the majority of WDs were observed in only the $K$ 
filter (only 28 of the P83 targets have measurements in all three 
bands, $JHK$). The presence of IR excess (indicating a possible 
cool companion) was deduced by comparison of the observed $K$ 
magnitude with that predicted by model calculations.  Out of the 
113 objects surveyed, only seven objects exhibited IR excess and 
could be readily deconvolved into a WD + red dwarf pair.  
Six additional objects are identified as ``anomalous composites,'' 
which showed moderate IR excess but could not be separated into a 
WD + red dwarf pair via model fits.  Our $K_{\rm s}$ band measurements 
agree well with those of P83 for almost all of the 49 targets in 
common between our studies.  Four of the binary candidates 
identified by P83 are located in the 2MASS 2IDR sky coverage.  
Three of these four (0034$-$211, 0429+176, and 1333+487) are listed 
in MS99 as binaries and are also found as binary candidates in 
our 2MASS data.  The fourth object, WD 1919+145, is listed in 
P83 as an ``anomalous composite'' with moderate IR excess; 
however, its 2MASS colors in our data place it in the locus of single WDs.

\citet{zuckerman92} expanded the survey by P83 to include 
$\sim 200$ WDs. They found likely cool companions within 
$6\arcsec$ of 21 WDs.  They do not provide their entire list 
of surveyed WDs (only the binary candidates), so we cannot 
perform a full comparison to our data.  Eight of their binary 
candidates overlap with our 2MASS sample and six of these 
(0710+741, 0752$-$146, 1026+002, 1123+189, 1210+464, and 2256+249) 
are also identified as binary candidates in our data.  
The remaining two are a resolved binary and an unresolved double degenerate.

Finally, GAN00 obtained $J$ and $K$ band observations of 49 
{\em Extreme Ultraviolet Explorer}-selected hot WDs.  
Ten of these WDs exhibit significant IR excess, five of which 
were previously known to be WD + red dwarf binaries.  
Thirty-four of the 49 WDs from GAN00 are located in the 
2MASS 2IDR sky coverage, but six of these are too faint and 
were not detected (another, WD0427+741J, was one of the ten 
found to have an IR excess, but it is near the faint detection 
limit for 2MASS, lacks formal photometric uncertainties, and is 
not included in our list of binary candidates).  Of the remaining 
27 objects that were detected by 2MASS, three are known unresolved 
binaries listed in MS99 (0148$-$255J, 1123+189 and 1631+781), 
while two more were identifed as binaries in other literature 
sources (0131$-$163 and 1711+667J) -- see Table \ref{t-bincans}.  
(These are the same five objects noted by GAN00 as previously 
known binaries.)

\section{Conclusions}

We have shown that the near-IR color-color diagram is an effective 
means of identifying candidate binary stars containing a WD and a 
low mass main sequence star.  The loci of single WDs and WD + red 
dwarf binaries occupy distinct regions of the near-IR color-color 
diagram.  Using our data from the 2MASS 2IDR, we recovered all 
known unresolved WD + red dwarf binaries located in the 2IDR sky 
coverage, and also identified nearly as many new candidate binaries 
(47 new candidates out of 95 total).  In addition, a handful of 
the known resolved binaries may actually be triple systems, in 
which the WD component is itself an unresolved WD + red dwarf 
binary.  We expect to be able to more than double again the number 
of candidate binaries using the forthcoming full sky data release 
from 2MASS.

Using observational near-IR data for WDs and M--L dwarfs, we have 
compared a sample of simulated WD + red dwarf binaries with our 
2MASS data.  The colors of the simulated binaries are dominated 
by the low mass companion through the late-M to early-L spectral 
types.  As the spectral type of the companion becomes progressively 
later, however, the colors of unresolved binaries become 
progressively bluer.  Binaries containing the lowest mass 
companions will be difficult to distinguish from the locus of 
single WDs in the near-IR color-color diagram.  We have identified 
an additional 15 WDs that may comprise such binaries.  It is 
encouraging that one of these has been found to be a close WD + red 
dwarf binary in which the companion has a mass of 
$\lesssim0.1M_{\odot}$ \citep{marsh96}.  In order to distinguish 
the WD + red dwarf binaries from single WDs for systems containing 
the lowest mass L dwarfs (and brown dwarfs), it is likely to be 
necessary to observe further into the infrared; for example, at 
the mid-IR wavelengths observable with the {\em Space Infrared 
Telescope Facility} (e.g., \citealt{ignace02}). 

\acknowledgments

We thank Roc Cutri (for a helpful discussion about the 2MASS colors 
of main sequence stars), 
Gus Muench-Nasrallah (for sharing his insight into initial 
mass functions), and Vandana Desai and Oliver Fraser (who 
helped locate WDs in 2MASS images in exchange for pizza).  
R.E.W. thanks the DeEtte McAuslan Stuart Scholarship Committee, 
the Boeing Company, and the National Merit Scholarship Corporation 
for their financial support.  The research described in this paper 
was carried out, in part, at the Jet Propulsion Laboratory, 
California Institute of Technology, and was sponsored by the 
National Aeronautics and Space Administration.  This publication 
makes use of data products from the 2 Micron All Sky Survey, 
which is a joint project of the University of Massachusetts and 
the Infrared Processing and Analysis Center/California Institute 
of Technology, funded by the National Aeronautics and Space 
Administration and the National Science Foundation. It also 
utilized NASA's Astrophysics Data System Abstract Service and 
the SIMBAD database operated by CDS, Strasbourg, France, as 
well as images from the Digitized Sky Survey, which was produced 
at the Space Telescope Science Institute under US Government 
grant NAG W-2166. (The images of these surveys are based on 
photographic data obtained using the Oschin Schmidt Telescope 
on Palomar Mountain and the UK Schmidt Telescope. The plates 
were processed into the present compressed digital form with 
the permission of these institutions.)

\clearpage

\clearpage

\begin{deluxetable}{lccccccc}
\tablecaption{White Dwarf + Low Mass Main Sequence Star Binary Candidates
\label{t-bincans}}
\tablehead{
\colhead{WD Number} & 
\colhead{Binary?} & 
\colhead{$J$} &
\colhead{$\sigma_J$} & 
\colhead{$H$} & 
\colhead{$\sigma_H$} & 
\colhead{$K_s$}  &
\colhead{$\sigma_{K_s} $}
}
\startdata
0023+388         & MS99\tablenotemark{a} & 13.807 & 0.029 & 13.250 & 0.033 & 12.947 & 0.039  \\
0034$-$211       & MS99\tablenotemark{b} & 11.431 & 0.032 & 10.911 & 0.029 & 10.636 & 0.030  \\
0102+210.2       & MS99\tablenotemark{c} & 16.701 & 0.114 & 16.221 & 0.161 & 15.574 & 0.206  \\
0116$-$231       & MS99                  & 14.602 & 0.036 & 14.064 & 0.043 & 13.803 & 0.054  \\
0130$-$196       & this work             & 14.785 & 0.037 & 14.270 & 0.040 & 14.023 & 0.058  \\
0131$-$163       & 1                     & 12.963 & 0.035 & 12.447 & 0.033 & 12.241 & 0.034  \\
0145$-$221       & this work             & 14.925 & 0.037 & 14.429 & 0.048 & 14.366 & 0.065  \\
0148$-$255J      & MS99                  & 12.472 & 0.026 & 11.882 & 0.034 & 11.593 & 0.029  \\
0145$-$174       & this work             & 15.177 & 0.052 & 14.646 & 0.069 & 14.330 & 0.074  \\
0205+133         & 2\tablenotemark{d}    & 12.797 & 0.030 & 12.196 & 0.029 & 11.950 & 0.025  \\
0208$-$153       & this work             & 12.621 & 0.029 & 12.081 & 0.026 & 11.778 & 0.030  \\
0219+282         & this work             & 16.067 & 0.078 & 15.587 & 0.103 & 15.293 & 0.146  \\
0252+209         & MS99                  & 16.482 & 0.110 & 16.044 & 0.147 & 16.097 & 0.322 \\
0257$-$005       & this work             & 16.773 & 0.139 & 16.344 & 0.211 & 15.526 & 0.204  \\
0303$-$007       & MS99                  & 13.165 & 0.027 & 12.625 & 0.025 & 12.410 & 0.030  \\
0309$-$275       & this work             & 13.523 & 0.034 & 12.881 & 0.030 & 12.738 & 0.033  \\
0324+738         & MS99\tablenotemark{e} & 11.719 & 0.029 & 11.086 & 0.024 & 10.822 & 0.026  \\
0347$-$137       & 1                     & 12.045 & 0.034 & 11.565 & 0.044 & 11.301 & 0.033  \\
0355+255         & MS99\tablenotemark{e} &  9.009 & 0.046 &  8.496 & 0.041 &  8.344 & 0.028  \\
0357+286J        & MS99                  &  9.845 & 0.046 &  9.260 & 0.033 &  9.048 & 0.032  \\
0357$-$233       & this work             & 15.054 & 0.048 & 14.587 & 0.058 & 14.266 & 0.073  \\
0408+158         & MS99                  & 10.777 & 0.037 & 10.199 & 0.037 &  9.926 & 0.030  \\
0413$-$077       & MS99                  &  6.738 & 0.019 &  6.279 & 0.041 &  5.966 & 0.047  \\
0429+176         & MS99                  & 10.755 & 0.032 & 10.115 & 0.035 &  9.927 & 0.036  \\
0430+136         & MS99                  & 13.550 & 0.033 & 12.877 & 0.039 & 12.655 & 0.041  \\
0458$-$662       & MS99                  & 13.431 & 0.032 & 12.686 & 0.026 & 12.511 & 0.035  \\
0628$-$020       & MS99\tablenotemark{f} & 10.704 & 0.027 & 10.150 & 0.029 &  9.838 & 0.026  \\
0752$-$146       & 3                     & 12.625 & 0.024 & 12.135 & 0.028 & 11.831 & 0.027  \\
0807+190         & 4                     & 15.790 & 0.082 & 15.210 & 0.118 & 15.229 & 0.160 \\
0812+478         & this work             & 14.578 & 0.038 & 14.149 & 0.048 & 13.849 & 0.067 \\
0825+367         & this work             & 14.077 & 0.045 & 13.507 & 0.043 & 13.318 & 0.053  \\
0851+190         & this work             & 15.514 & 0.056 & 15.029 & 0.079 & 14.597 & 0.077  \\
0904+391         & this work             & 15.436 & 0.061 & 14.893 & 0.077 & 14.592 & 0.085  \\
0908+226         & MS99                  & 15.184 & 0.048 & 14.473 & 0.042 & 14.350 & 0.061  \\
0915+201         & this work             & 15.712 & 0.064 & 15.148 & 0.076 & 14.824 & 0.083  \\
0937$-$095       & MS99                  & 13.797 & 0.032 & 13.210 & 0.027 & 13.058 & 0.038 \\
0950+139         & 5                     & 16.430 & 0.114 & 15.555 & 0.144 & 15.350 & 0.151  \\
0954+134         & this work             & 15.561 & 0.069 & 14.838 & 0.087 & 14.548 & 0.089  \\
1001+203         & MS99                  & 12.642 & 0.040 & 12.020 & 0.032 & 11.756 & 0.037  \\
1013$-$050       & MS99                  & 10.635 & 0.028 &  9.985 & 0.029 &  9.775 & 0.027  \\
1026+002         & MS99                  & 11.771 & 0.032 & 11.218 & 0.032 & 10.916 & 0.027  \\
1037+512         & this work             & 13.804 & 0.031 & 13.235 & 0.028 & 12.973 & 0.029  \\
1054+305         & MS99                  & 11.888 & 0.031 & 11.168 & 0.053 & 10.982 & 0.023  \\
1054+419         & MS99                  &  9.473 & 0.032 &  8.863 & 0.031 &  8.619 & 0.033  \\
1106+316         & this work             & 15.116 & 0.048 & 14.543 & 0.053 & 14.474 & 0.102  \\
1106$-$211       & this work             & 14.673 & 0.040 & 13.910 & 0.047 & 13.814 & 0.058  \\
1108+325         & this work             & 15.785 & 0.072 & 15.204 & 0.084 & 15.187 & 0.181  \\
1123+189         & MS99                  & 12.777 & 0.038 & 12.232 & 0.035 & 11.999 & 0.025  \\
1133+358         & MS99                  & 11.631 & 0.036 & 11.101 & 0.054 & 10.780 & 0.037  \\
1136+667         & MS99                  & 12.369 & 0.030 & 11.749 & 0.034 & 11.615 & 0.038  \\
1156+129         & this work             & 14.702 & 0.045 & 14.104 & 0.044 & 13.885 & 0.051  \\
1156+132         & this work             & 16.886 & 0.148 & 16.224 & 0.189 & 15.939 & 0.209  \\
1201+437         & MS99                  & 15.410 & 0.050 & 14.829 & 0.061 & 13.777 & 0.043  \\
1210+464         & MS99                  & 12.076 & 0.029 & 11.414 & 0.030 & 11.167 & 0.027  \\
1211$-$169       & this work             &  7.945 & 0.025 &  7.340 & 0.031 &  7.190 & 0.042  \\
1214+032         & MS99                  &  9.220 & 0.030 &  8.648 & 0.030 &  8.412 & 0.027  \\
1218+497         & this work             & 14.579 & 0.041 & 13.977 & 0.042 & 13.830 & 0.063  \\
1224+309         & 6                     & 15.122 & 0.058 & 14.687 & 0.069 & 14.308 & 0.084  \\
1229+290         & this work             & 15.888 & 0.084 & 15.210 & 0.102 & 14.561 & 0.099  \\
1236$-$004       & this work             & 16.386 & 0.113 & 15.871 & 0.145 & 15.558 & 0.236  \\
1247$-$176       & 7                     & 13.536 & 0.034 & 12.892 & 0.031 & 12.601 & 0.034  \\
1302+317         & this work             & 15.837 & 0.070 & 15.295 & 0.091 & 15.113 & 0.130  \\
1307$-$141       & this work             & 13.869 & 0.033 & 13.229 & 0.040 & 13.019 & 0.044  \\
1330+793         & MS99                  & 12.482 & 0.028 & 11.863 & 0.026 & 11.697 & 0.028  \\
1333+487         & MS99                  & 11.829 & 0.024 & 11.260 & 0.027 & 10.960 & 0.033  \\
1339+346         & this work             & 14.095 & 0.041 & 13.695 & 0.043 & 13.572 & 0.047  \\
1412$-$049       & this work             & 13.726 & 0.029 & 13.107 & 0.034 & 12.975 & 0.038  \\
1431+257         & this work             & 16.260 & 0.100 & 15.542 & 0.101 & 15.714 & 0.224 \\
1435+370         & this work             & 13.467 & 0.030 & 12.954 & 0.039 & 12.760 & 0.032  \\
1436$-$216       & this work             & 13.326 & 0.029 & 12.757 & 0.030 & 12.488 & 0.035  \\
1443+336         & this work             & 14.265 & 0.034 & 13.697 & 0.042 & 13.509 & 0.047  \\
1458+171         & this work             & 14.676 & 0.038 & 14.174 & 0.050 & 13.743 & 0.057  \\
1502+349         & this work             & 15.208 & 0.050 & 14.759 & 0.071 & 14.300 & 0.072  \\
1504+546         & 8                     & 13.854 & 0.029 & 13.250 & 0.034 & 13.003 & 0.033  \\
1517+502         & MS99\tablenotemark{g} & 15.553 & 0.064 & 14.744 & 0.075 & 14.133 & 0.072  \\
1522+508         & this work             & 14.737 & 0.041 & 14.196 & 0.049 & 13.921 & 0.057  \\
1527+450         & this work             & 16.212 & 0.090 & 15.784 & 0.117 & 15.350 & 0.205  \\
1558+616         & MS99                  & 14.207 & 0.036 & 13.609 & 0.047 & 13.349 & 0.047  \\
1603+125         & this work             & 13.544 & 0.033 & 13.088 & 0.036 & 12.986 & 0.031  \\
1606+181         & this work             & 14.778 & 0.039 & 14.142 & 0.055 & 13.910 & 0.054  \\
1610+383         & this work\tablenotemark{h} & 14.404 & 0.045 & 13.750 & 0.046 & 13.560 & 0.060  \\
1619+525         & this work             & 14.178 & 0.035 & 13.619 & 0.040 & 13.421 & 0.042  \\
1619+414         & MS99                  & 13.918 & 0.033 & 13.272 & 0.040 & 13.009 & 0.041  \\
1622+323         & MS99                  & 14.644 & 0.040 & 13.991 & 0.040 & 13.796 & 0.050  \\
1631+781         & MS99                  & 10.998 & 0.031 & 10.381 & 0.031 & 10.154 & 0.026  \\
1643+143         & 1                     & 12.766 & 0.035 & 12.096 & 0.055 & 11.955 & 0.028  \\
1654+160         & 9                     & 13.066 & 0.045 & 12.402 & 0.059 & 12.145 & 0.042  \\
1711+667J        & 10                    & 15.088 & 0.045 & 14.430 & 0.059 & 14.213 & 0.086  \\
1717$-$345       & 11                    & 12.864 & 0.027 & 12.244 & 0.052 & 12.008 & 0.046  \\
2133+463         & MS99                  & 11.341 & 0.026 & 10.747 & 0.028 & 10.459 & 0.032  \\
2151$-$015       & MS99                  & 12.478 & 0.030 & 11.787 & 0.025 & 11.443 & 0.031  \\
2256+249         & MS99                  & 11.663 & 0.039 & 11.204 & 0.041 & 10.892 & 0.033 \\
2317+268         & this work             & 14.614 & 0.032 & 14.067 & 0.040 & 13.769 & 0.050  \\
2323+256         & this work             & 15.815 & 0.077 & 15.404 & 0.133 & 15.385 & 0.164  \\
2326$-$224       & this work             & 12.649 & 0.028 & 12.031 & 0.039 & 11.753 & 0.030  
\enddata
\tablerefs{
(1) \citet{szb96}; 
(2) \citet{greenstein86b}; 
(3) \citet{schultz93}; 
(4) \citet{gizis97}; 
(5) \citet{fulbright93}; 
(6) \citet{orosz99}; 
(7) \citet{koester01}; 
(8) \citet{stepanian01};
(9) \citet{zuckerman92}; 
(10) \citet{fkb97};
(11) \citet{reid88}.
}
\tablenotetext{a}{WD0023+388 is a known triple system composed of a close WD + red dwarf pair 
with a wide red dwarf companion \citep{reid96}.}
\tablenotetext{b}{WD0034$-$211 is classified as a close double degenerate binary in MS99. 
\citet{brag90} reclassified it as a WD + red dwarf binary, which is supported by its 2MASS colors.}
\tablenotetext{c}{WD0102+210.2 is classified as one component of a wide double degenerate binary 
\citep{sion91}, but our 2MASS colors suggest that it  may be a close WD + red dwarf binary also.}
\tablenotetext{d}{WD0205+133 may be a sdOB + red dwarf binary, instead of a WD + red dwarf binary \citep{allard94}.}
\tablenotetext{e}{This object is classified as the WD component of a wide (resolved) WD + red 
dwarf binary, but our 2MASS colors suggest that the WD may also be a close WD + red dwarf binary.}
\tablenotetext{f}{The binary separation of WD0628$-$020 ($4\arcsec$) is near the 2MASS resolution
limit; there is only one entry in the 2IDR point source catalog, but these magnitudes may be for 
the red dwarf component only.}
\tablenotetext{g}{The companion of WD1517+502 is a dwarf carbon star \citep{liebert94}.}
\tablenotetext{h}{WD1610+383 is barely resolved on the DSS images as a common proper motion pair 
with red and blue components at a separation of $\approx4$ arcsec.  Blinking of the DSS and 2MASS
images suggests that only the red component is detected by 2MASS.  Its near-IR colors are 
consistent with an early-M spectral type.}
\end{deluxetable}

\begin{deluxetable}{lccccccc}
\tablecaption{Tentative WD + Low Mass Main Sequence Star Binary Candidates
\label{t-tentbincans}}
\tablehead{
\colhead{WD Number} & 
\colhead{Binary?} & 
\colhead{$J$} &
\colhead{$\sigma_J$} & 
\colhead{$H$} & 
\colhead{$\sigma_H$} & 
\colhead{$K_s$}  &
\colhead{$\sigma_{K_s} $}
}
\startdata
0023$-$109 & MS99\tablenotemark{a} & 16.042 & 0.081 & 15.852 & 0.172 & 15.608 & 0.237  \\
0029$-$032 & this work             & 15.660 & 0.055 & 15.380 & 0.089 & 15.172 & 0.151  \\
0518+333   & MS99\tablenotemark{a} & 15.439 & 0.082 & 15.184 & 0.103 & 14.912 & 0.105  \\
0710+741   & 1                     & 14.708 & 0.035 & 14.425 & 0.059 & 14.155 & 0.070  \\
0816+387   & MS99\tablenotemark{a} & 16.036 & 0.106 & 15.765 & 0.196 & 15.549 & 0.229  \\
0942+236.1 & MS99\tablenotemark{a} & 16.675 & 0.115 & 16.292 & 0.178 & 16.059 & 0.256  \\
1008+382   & this work             & 16.883 & 0.154 & 16.646 & 0.263 & 16.290 & 0.300  \\
1015$-$173 & this work             & 15.241 & 0.049 & 14.863 & 0.069 & 14.569 & 0.107  \\
1247+550   & this work\tablenotemark{b} & 15.782 & 0.071 & 15.618 & 0.135 & 15.293 & 0.469  \\
1434+289   & this work             & 16.544 & 0.120 & 16.334 & 0.206 & 15.946 & 0.301  \\
1639+153   & this work             & 15.065 & 0.049 & 14.960 & 0.069 & 14.595 & 0.132  \\
2211+372   & this work             & 16.278 & 0.106 & 16.057 & 0.195 & 15.736 & 0.246  \\
2257+162   & this work             & 15.401 & 0.062 & 15.045 & 0.074 & 14.674 & 0.110  \\
2336$-$187 & this work             & 15.057 & 0.041 & 14.923 & 0.069 & 14.694 & 0.097  \\
2349$-$283 & this work             & 16.119 & 0.090 & 15.863 & 0.178 & 15.549 & 0.222  
\enddata
\tablerefs{
(1) \citet{marsh96}.
}
\tablenotetext{a}{This object is classified as the WD component of a wide (resolved) binary, but 
our 2MASS colors suggest that the WD may also be a close WD + red dwarf binary.}
\tablenotetext{b}{MS99 note that WD1247+550 is ``possibly the coolest known degenerate star.''}
\end{deluxetable}

\end{document}